\documentclass[twocolumn,twoside,slac,floatfix]{revtex4}
\usepackage{graphicx}
\usepackage{fancyhdr}
\begin{document}

\title{Setting confidence intervals in coincidence search analysis }

%

\author{L. Baggio}
\affiliation{INFN and Trento Univ., Povo, TN 38050, Italy}

\author{G.A. Prodi}
\affiliation{INFN and Trento Univ., Povo, TN 38050, Italy}

\begin{abstract}
The main technique that has been used to estimate the rate of gravitational wave (gw) bursts is to search for coincidence among times of arrival of candidate events in different detectors. Coincidences are modeled as a (possibly non-stationary) random time series background  with gw events embedded in it, at random times but constant average rate. It is critical to test whether the statistics of the coincidence counts is Poisson, because the counts in a single detector often are not. At some point a number of parameters are tuned to increase the chance of detection by reducing the expected background: source direction, epoch vetoes based on sensitivity, goodness-of-fit thresholds, etc. Therefore, the significance of the confidence intervals itself has to be renormalized. This review is an insight of the state-of-the-art methods employed in the recent search performed by the International Gravitational Event Collaboration for the worldwide network of resonant bar detectors.
\end{abstract}

\maketitle

\thispagestyle{fancy}


\section{Introduction}
When a detector is pushed to its limits in order to reveal faint sources, every slight deviation of noise models from ideality can severely jeopardize the robustness of a detection claim. In fact, when the signal-to-noise ratio (SNR) is low, most goodness-of-the-fit tests have poor discrimination power. On the other hand, in the long run, the outliers add up and constitute a background which can be much larger than the isolated signals possibly present in the data.

Working with a network of detectors optimized for coincidence analysis allows to reduce the background and --most of all-- to estimate reliably the background itself, which is essential to set reliable upper limits.

A gravitational wave (gw) resonant detector is built around a mechanically isolated massive resonant body. Cylindric 3m-long 2.3 ton aluminum alloy bars have been until now a widely adopted solution. Any planar (transverse) gravitational wave impinging on the bar with an angle $\theta$ relative to its axis excites the longitudinal mechanical mode, with amplitude proportional to $\sin^2\theta$.
With respect to
burst signals,
the presently working resonant detectors are sensitive in a narrow ($\sim 1-10Hz$) frequency range near the resonance ($\sim 900Hz$).~\cite{IGEC-PRD}

A candidate event is defined as the output of an automated max-hold algorithm based on two adaptive thresholds: one on the SNR of the peak amplitude (it has to be great enough to be identified without ambiguity, i.e. low timing error) and one on the minimum delay between consecutive events (in order to generate independent events it must be greater than a few times the autocorrelation of the processed data). Even with no outliers, this algorithm would produce random accidental events as samples of the extreme distribution for an (almost) Gaussian stochastic process.

The International Gravitational Events Collaboration (IGEC)~\cite{IGEC-PRL, IGEC-PRD, IGEC-WWW} was founded in order to take up the task of assessing the detection of gw's from the candidate event lists compiled by the single detectors.
The only requirement for member groups has been that the exchanged information should include:

i) event amplitudes and times of arrival (along with their estimated errors)

ii) minimal detectable amplitude --i.e. the sensitivity threshold of the detector-- defined by requirement of unbiasedness of amplitude estimates and unambiguous timing.

The IGEC analysis is based on time coincidence search, and in the first 4 year run (1997-2000) the five detectors of the collaboration were purposely aligned to be as parallel as possible, in order to maximize the efficiency of the network. The analysis as it was recently performed is still not optimal in many respects. Moreover, because the gw source amplitude distribution and polarization are unknown, the detection efficiency is not completely determined. However, with respect to past and recent proposals, this analysis improves the control of probability of false dismissal of candidate gw signals and provides the detailed computation of the probability of accidental detection.

In the IGEC analysis many selections and tests are applied to the data, in order to enhance the chances of gw detection as a function of the amplitude and direction of target gw signals. The selections may enhance accidental detections as well, therefore a record of all the attempts has to be compiled. When a complete account is given for all the operations on the data, and assuming that their statistics is known, the probability that any of the observed results is due to chance can be well accounted for within the frequentist framework.

\section{Data cuts and coincidence search}
\begin{figure*}[ht]
\centering
\includegraphics[width=120mm]{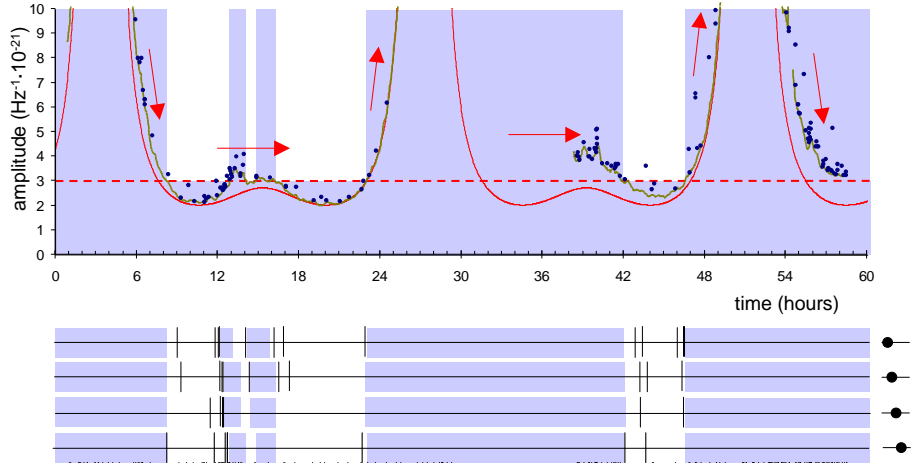}
\caption{\label{fig:resampling} (\textit{above}) Example of data selection in a time span of a few hours. The amplitude is given in terms of spectral density of the gw strain at a frequency about $900Hz$, assuming one specific direction (galactic center), and neglecting polarization. From the original event time series (\textit{dots}) after angular sensitivity modulation (\textit{solid smooth curve}) only those are retained whose amplitude is above a fixed common absolute threshold (\textit{dashed line}). Correspondingly, the periods when the local detector threshold (\textit{solid crispy curve}) is above the common threshold are removed from the observation time (\textit{vertical solid shadows}). This generates the ``on-source" time series (\textit{below, top row}). To obtain resampled (and ``off-source") event selections (\textit{below, under the first row}), the local time coordinate at the detector site is shifted by a proper amount (\textit{arrows}).
It is worth noticing that the background event density drops exponentially toward greater amplitudes. The density of event amplitude \textit{relative to the local threshold} is more or less the same at all times, but \textit{relative to the fixed common threshold} it is highly nonstationary. In fact almost all events are cut out by the selection mechanism except for when the local threshold approaches closely the common threshold from below --i.e. near the edges of the live time spans. The angular sensitivity modulation (which is similar in parallel detectors) enhances this mechanism of artificial clustering, and it generates a remarkable cross-correlation of event rates between detectors. The described resampling procedure preserves  the correlation pattern.}
\end{figure*}

Hereafter the focus will be a source located in the direction of the galactic center, as it is likely that the present sensitivity of bar detectors limits the observation range to sources within the Milky Way. The times of arrival are supposed to be already corrected for the light travel time delay for detectors at different positions. Moreover, as discussed in Fig.~\ref{fig:resampling}, the measured amplitude of events has been corrected for the angular sensitivity factor.

A (twofold) \textit{coincidence} is defined when the time of arrival $t_i$ and $t_j$ of two events from different detectors satisfies the inequality
\begin{equation}
\label{eq:coincidence}
|t_i-t_j|<k \sqrt{\sigma_i^2+\sigma_j^2}
\end{equation}
where $\sigma_i$ and $\sigma_j$ are the standard deviation of the time error, and $k$ depends on the target false dismissal. The timing error is not Gaussian,  and its standard deviation is strongly dependent on the signal-to-noise ratio of the event amplitude (it ranges from a second down to milliseconds; see for instance Fig.~1 in Ref.~\cite{IGEC-toolbox}). A conservative value for $k$ is given by the Bienaym\`e-Tchebycheff inequality: the probability that the absolute value of a zero mean random variable is greater than $k$ times its standard deviation $\sigma$ is $P(|x|>k \sigma)\le k^{-2}$. For instance, $k\sim 4.5$ guarantees a false dismissal less than 5\%.

In general, an $M$-fold coincidence is defined as the simultaneous coincidence in the $M(M-1)$ distinct couples out of M detectors. In this case, for a target false dismissal probability $P_T$, one has to set $k=(1-(1-P_T)^{2/[M(M-1)]})^{-1/2}$.
As for the rate of accidental coincidences, it is proportional to $k^{M-1}$ and to the rate of events in each individual detector\cite{IGEC-toolbox}.

The IGEC adopted the following data selection scheme (see Fig.~\ref{fig:resampling}):

\noindent i) fix a common (absolute) threshold $A_{th}$;

\noindent ii) cut the time spans when the minimal detectable amplitude of each detector was greater than $A_{th}$;

\noindent iii) within these periods, include only those events with amplitude greater than $A_{th}$.

We investigated different results from many values of $A_{th}$,
and consequently
we 
accounted for the increased probability of false alarm\footnote{Actually, in Refs.~\cite{IGEC-PRL,IGEC-PRD} it is a common practice to perform the analysis separately on disjoint subsets of the data, each one pertaining to a different configuration of the network --i.e. different combinations of detectors in common operation. Eventually, the data are re-aggregated per equal amplitude threshold.
} (see Sec.~\ref{sec:confidence}).


\section{Background estimate}

The IGEC uses resampling methods to estimate the rate of uncorrelated background coincidences. Approximately randomized samples of the coincidence counts can be obtained by rigidly shifting the times of arrival of the original event time series of individual detectors relative to each other. With this new data set, the whole analysis is repeated: amplitude modulation, data selection and coincidence search.

The choice of a rigid time shift instead of reshuffling or swapping is due to the presence of structures in the autocorrelation of the single detector event time series, with characteristic timescales from a few seconds to one minute (see Fig.~8 in Ref.~\cite{IGEC-PRD}) --i.e. the time series are not Poisson. Moreover, the angular modulation and the common amplitude thresholding applied to the data conspire to produce further event clustering (see Fig.~\ref{fig:resampling}). A rigid time shift guarantees that all these structures are not smoothed out when generating resampled counts.

In order to obtain \textit{independent} resampled counts, the time series were always shifted more than the maximum \textit{time window} (i.e. the right side of Eq.~\ref{eq:coincidence}) ever used (in practice, few seconds).

To test that the resampled counts come from the same statistic, and that the latter is Poisson, the histograms of coincidence counts were fitted with a Poisson probability density profile. The one-tail $\chi^2$-test has been performed on every network configuration (provided that at least one degree of freedom was available), and the histogram of the computed p-levels was in agreement with uniform density, which is the expected one if the model of the background statistic is good.

Strictly speaking, what has been verified is just the coherence of the resampling approximation --all resampled counts due to the same statistic. This result holds up to timescales of the order of one hour, which translates in a few thousands of independent resampled coincidence counts. The statistical error for the resampled background rate is then about 3\%.

However, in order to conclude that the resampled statistics is also identical to the statistic of the unshifted original data, one has to be confident that no source of correlated background events exists. This ansatz is assumed without proof.


\section{Confidence intervals}
\label{sec:confidence}

The results of IGEC search are frequentist, i.e. the quoted confidence level or coverage are meant to be --at least conservatively-- the probability that the confidence interval contains the true value. This approach is also unified in that it prescribes how to set a confidence interval automatically leading to a claim of detection or an upper limit. The construction of the confidence belt however does not proceed \textit{\`a la} Feldman and Cousins~\cite{FC}, or proposed modifications, where the coverage is kept as fixed as possible for any source strength. Instead, the confidence interval bounds are independently derived from the likelihood function. This inevitably leads to variable coverage, and we shall show briefly how the minimum of the coverage is related to the integral of the likelihood.\cite{Porter,RW2k}

Let $N_c = N_b  + N_\Lambda$ where $N_c$ are the counted coincidences, $N_b$ those due to background,  $N_\Lambda$ those due to a hypothetical flux of gw's with mean rate $\Lambda$; let also $\mu _c$, $\mu _b$ and $\mu _\Lambda$ be their mean values, respectively. The probability density function under the hypothesis of a Poisson statistic for both $N_b$ and $N_\Lambda$ is
\begin{equation}
f(N_c ; \mu _\Lambda ,\mu _b  ) = \frac{{e^{ - \left( {\mu _b  + \mu _\Lambda } \right)} }}
{{N_c !}}\left( { \mu _b  + \mu _\Lambda  } \right)^{N_c }
\end{equation}
and the likelihood function is defined as usual as $\ell (\mu_\Lambda  ;N_c ,\mu_b ) \equiv f(N_c ;\mu_\Lambda ,\mu_b )$. Let $I$ be a parameter from 0 to 1; one has to solve for $0 \leq N_{\inf }<N_{\sup }$ the equations

\begin{equation}
\left\{ \begin{array}{l}
 \ell (n_{\inf } ;N_c , \mu_b) = \ell (N_{\sup } ;N_c ,\mu_b) \\
 N_{\inf } = \max ( n_{\inf }, 0)\\
 I = \left[ {\int_0^\infty  {\ell (\mu ;N_c ,\mu_b)d\mu  } } \right]^{ - 1} \int_{N_{\inf } }^{N_{\sup } } {\ell (\nu  ;N_c ,\mu_b)d\nu }  \\
 \end{array} \right.
\end{equation}

\begin{figure}[t]
\centering
\includegraphics[width=78mm]{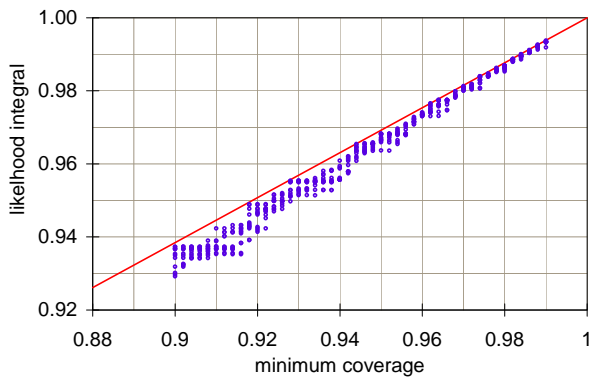}
\caption{\label{fig:equivalence} Integral of Poisson likelihood $I$ vs minimum coverage of $\mu_\Lambda$, for various choices of the background: $\mu_b\in \{0.01, 0.02, 0.05, 0.1, ..., 20, 50\}$. For any chosen value of $I$ and $\mu_b$, each dot was obtained by scanning a range of source rates, computing the coverage at each one and then taking the minimum. The relation between $I$ and $\mu_b$ depends weakly on the background and is approximately linear.}
\end{figure}

The interval for $\mu_{\Lambda }$, delimited by $N_{\inf }$ and $N_{\sup }$, maximizes the integral of the likelihood in the physical domain $\mu_{\Lambda }\ge 0$, hence it belongs to a set which can be derived by a Bayesian procedure assuming constant prior for $\mu _\Lambda \ge0$. However, we would give to this intervals frequentist interpretation, by computing the coverage
\begin{equation}
C(\mu_\Lambda  ) \equiv \sum\limits_{N_c |N_{\inf }  < \mu_\Lambda   < N_{\sup } } {f(N_c ;\mu_\Lambda ,\mu_b)}
\end{equation}
The sum runs over the possible outcomes $N_c$ for which the interval $N_{\inf }-N_{\sup }$ covers the given value of $\mu_\Lambda$.
The coverage depends on $\mu_\Lambda$, hence to be conservative we refer to the coverage $C_{min}$ at the least covered value of $\mu_\Lambda$: $C_{min} \equiv \min\limits_{ \mu > 0} C(\mu)$. In Fig.~\ref{fig:equivalence} the relation between $I$ and $C_{min}$ has been computed numerically, for various values of $\mu_b$.

The choice of this procedure for IGEC analysis was first announced in Ref.~\cite{IGEC-toolbox}, but in that paper the effective coverage of the procedure is not pointed out. Ref.~\cite{RW2k} describes the same approach, but it also suggests \textit{ad hoc} modifications to improve the relation between the coverage and the integral of the likelihood. We think that this modification could jeopardize robustness, in particular when errors in the estimated background are not completely negligible. Fig.~3 in Ref.~\cite{IGEC-toolbox} shows a sample confidence belt originating from this method, and the uncertainty on confidence interval bounds due to uncertainty on $N_b$.

When $N_{\inf }$ and $N_{\sup }$ have been computed, one divides them by the length of the selected observation time, obtaining the bounds $\Lambda_{\inf }$ and $\Lambda_{\sup }$ on the flux of gw bursts whose \textit{measured amplitude} is above the common threshold. This limit is obviously cumulative, as lower flux is expected at higher thresholds. The details on how to unfold the results in terms of the \textit{true amplitude} go beyond the scope of this paper.

Many selection thresholds were tried, and all of these selections happened to be independent, as we shall say in a moment. As a result, the coverage of a single confidence interval does not tell the whole story. On one hand, a confidence interval set at lower selection threshold reinforces the confidence of the exclusion region resulting from a higher threshold where the exclusion regions overlap. On the other hand, even if there are actually no true gw events, after many trials a confidence interval excluding $\Lambda=0$ will eventually come out accidentally, as the coverage probability for $\mu_\Lambda=0$ --i.e. ${C(0)}$-- is not 1. This would lead to falsely reject the null hypothesis.

In order to compute correctly the probability of false claim (defined as \textit{at least} one interval not containing $\Lambda=0$) two methods were investigated.

First, if one assumes that the measures coming from different selections are independent random variables, then the probability of an accidental claim in case the null hypothesis is true is given by $1 - \prod\nolimits_i {C^{(i)} (0)}$, where the index $i$ runs over all different data selections. Notice that in the Poisson case $C^{(i)}(0) > C_{min}^{(i)}$ always, and $C^{(i)}(0)$ \textit{depends} on the background $\mu_b^{(i)}$.

Another method, which requires less assumptions, consists in resampling the entire list of results using the same randomizing procedure described above. In other words, the confidence intervals are computed on time-shifted data, for which we do not expect any genuine disagreement with the null result. From the resampled population of the would-be claims one can compute directly the chance of false alarm.

The two methods gave consistent results, which is in turn an evidence for independence of the different data selections\footnote{Of course, this depends on the coarseness of the chosen stepping for the common amplitude threshold. With finer steps one would expect correlation between nearby selections.}.

In this way the interpretation of the measure has two layers. We start from the bare confidence intervals, and count the ones which individually would deny the null hypothesis. Then we compare this number with the expected false claims. In the end, we get a confidence interval on the number of \textit{true} claims --if it includes zero, then we assess that no significant deviation from the null hypothesis was observed.


As a final remark, one should be aware that the number of papers quoting ``95\%" results \textit{just in the gw search field} has grown such that it would not be surprising to find a positive result among them by chance. If a sequence of negative results has just been observed, the first false positive is coming from the last --supposedly better-- experiment. It is really tempting to forget about the many previous null attempts (even easier if they were not published). However, a similar configuration can be just accidental (and much more than 5\% likely). This should be kept in mind when hurrying to claim the first non-null result in a series of many independent attempts --it is perhaps advisable to wait until it has been confirmed by successive experiments. Another solution would be to quote ``99\%" (or higher) confidence results, which give lower probability of a false claim. But this is not always possible because of limitations in the degree of accuracy of the noise models (in our case, it would require a more powerful test on the tails of the density function of $N_c$).

\begin{acknowledgments}
The authors wish to acknowledge the useful discussions within IGEC, and with J. Linnemann.

\end{acknowledgments}


\end{document}